\documentclass[preprint,proceedings]{rmaa}
\usepackage{paralist}
\usepackage{psfrag,color}

\newcommand{\mdotth}{\dot{M}_{-3}}
\newcommand{\mns}{M_{1.4}}
\newcommand{\begit}{\begin{itemize}}
\newcommand{\enit}{\end{itemize}}
\newcommand{\begen}{\begin{enumerate}}
\newcommand{\enen}{\end{enumerate}}

\newcommand{\beq}{\begin{equation}}
\newcommand{\eeq}{\end{equation}}

\newcommand{\beqa}{\begin{eqnarray}} 
\newcommand{\eeqa}{\end{eqnarray}}

\SetYear{2005}
\SetConfTitle{Triggering Relativistic Jets}

\title{Assessing Millisecond Proto-Magnetars as \\ GRB Central Engines} 

\author{Todd A.~Thompson\altaffilmark{1,2}}
\altaffiltext{1}{Lyman Spitzer Jr.~Fellow}
\altaffiltext{2}{Department of Astrophysical Sciences, Princeton University}

\shortauthor{Todd A.~Thompson}
\shorttitle{RevMexAA}

\fulladdresses{
\item Todd A.~Thompson: Department of Astrophysical Sciences, Peyton Hall-Ivy Lane,
Princeton University, Princeton, NJ 08544 (\email{thomp@astro.princeton.edu}).}

\listofauthors{Todd A.~Thompson}

\indexauthor{Thompson, Todd A.}

\abstract{Magnetars are a sizable subclass of the 
neutron star census.  Their very high magnetic field strengths
are thought to be a consequence of rapid (millisecond) rotation
at birth in a successful core-collapse supernova.  In their first
tens of seconds of existence, magnetars transition from hot,
extended ``proto-''magnetars to the cooled and magnetically-dominated
objects we identify $\sim10^4$ years later as Soft Gamma-ray Repeaters (SGRs)
and Anamolous X-ray Pulsars (AXPs).  Millisecond proto-magnetar winds during 
this cooling phase likewise transition from non-relativistic and thermally-driven to 
magneto-centrifugally-driven, and finally to relativistic and Poynting-flux dominated.
Here we review the basic considerations associated with that transition. 
In particular, we discuss the spindown of millisecond proto-magnetars throughout
the Kelvin-Helmholtz cooling epoch.  
Because of their large reservoir of rotational energy, their 
association with supernovae, and the fact that their winds are 
expected to become highly relativistic in the seconds after their 
birth, proto-magnetars have been suggested as the central engine of 
long-duration gamma ray bursts. 
We discuss some of the issues and outstanding questions
in assessing them as such.}

\addkeyword{Stars: Neutron, Magnetars}
\addkeyword{Stars: Stellar Winds, Outflows}
\addkeyword{Supernovae: General}
\addkeyword{Magnetohydrodynamics}
\addkeyword{Gamma Ray Bursts}

\begin{document}

\maketitle

\section{Introduction}

Core-collapse supernovae (SNe) leave behind hot 
proto-neutron stars that cool on the Kelvin-Helmholtz 
timescale ($\tau_{\rm KH}\approx10-100$ s) by
radiating their gravitational binding energy ($\sim10^{53}$ ergs) in neutrinos
(Burrows \& Lattimer 1986; Pons et al.~1999).  
A fraction of these neutrinos deposit their energy in the tenuous 
and extended atmosphere of the PNS.\footnote{The charged-current
interactions $\nu_e n\leftrightarrow p e^-$ and $\bar{\nu}_e p\leftrightarrow n e^+$
generally dominate heating and cooling.  For outflows with high entropy per baryon
$e^+e^-\leftrightarrow\nu\bar{\nu}$ and inelastic neutrino-electron/positron
scattering also contribute.}  In the standard picture, net neutrino heating drives a thermal 
wind that emerges into the post-supernova shock environment,
blowing a wind-driven bubble into the exploding and expanding supernova cavity 
(Woosley et al.~1994; Burrows, Hayes, \& Fryxell 1995).  For most massive stellar
progenitors with extended hydrogen envelopes (Type-II), the cooling phase is over well
before shock breakout at the progenitor's surface ($\sim$ 1 hour after collapse).  
For compact Type-Ibc supernovae, the supernova shockwave traverses the progenitor 
on a timescale  comparable to $\tau_{\rm KH}$.

For typical non-rotating non-magnetic (NRNM) neutron stars the wind/cooling
epoch is in some sense a small perturbation to the supernova event as a whole. 
Depending on how one defines the start of the wind phase, 
the total wind kinetic energy over $\tau_{\rm KH}$ is 
of order $10^{48}-10^{49}$ ergs, small on  on the scale of the
supernova explosion energy, $E_{\rm SN}\sim10^{51}$ ergs.\footnote{However, 
see Burrows et al.~(1995) and Scheck et al.~(2006)
for discussion of how the very early wind is tied to the supernova explosion itself.}
In addition, the total amount 
of mass ejected is a mere $\sim$$10^{-4}-10^{-3}$ M$_\odot$,  a minor addition to the
few solar masses ejected from the massive stellar progenitor during the 
explosion.
Finally, because of inefficient neutrino heating, the asymptotic wind speed does not 
exceed $\sim$0.1$c$ (Duncan, Shapiro, \& Wasserman 1986, Qian \& Woosley 1996).  

The primary focus of many previous efforts to understand proto-neutron star
winds --- particularly in the NRNM limit --- has been to assess these outflows
as the astrophysical site for production of the $r$-process nuclides
(Woosley et al.~1994; Takahashi et al.~1994; Qian \& Woosley 1996; 
Cardall \& Fuller 1997; Otsuki et al.~2000; 
Sumiyoshi et al.~2000; Wanajo et al.~2000; 
Thompson et al.~2001; for a review, see Thompson 2003a).

\subsection{Rotation \& Magnetic Fields}

Simple estimates imply that $\gtrsim$10\% of all supernovae produce 
magnetars, a class of young neutron stars (the Anamolous X-ray Pulsars 
and Soft Gamma-ray Repeaters)  with inferred large-scale
surface magnetic dipole fields in the range $10^{14}-10^{15}$ G
(Duncan \& Thompson 1992; Thompson \& Duncan 1993; Kouveliotou et al.~1999;
see also Woods \& Thompson 2004; Kaspi \& Helfand 2002).
The high magnetic fields of magnetars 
may result from the collapse of ultra-magnetized 
iron cores in massive stars or white dwarfs in accretion-induced
collapse, or they may be generated by rapid rotation 
and efficient dynamo action during the Kelvin-Helmholtz cooling epoch
of proto-neutron stars
(see below, Duncan \& Thompson 1992). In this latter scenario,
it is millisecond spin periods at birth that are thought to 
distinguish magnetars from normal neutron stars, with their 
characteristically lower field strengths.

It is interesting to consider how the standard picture of NRNM
neutrino-driven protoneutron star winds is modified by the 
presence of a strong large-scale magnetic field and rapid rotation.  
To do so, we first quantify the meaning of ``strong magnetic field'' 
and ``rapid rotation'' in the proto-neutron star context.
For this purpose, it useful to consider the physical conditions of the 
proto-neutron star atmosphere just after birth.  

Assuming a successful 
explosion occurs, the total neutrino luminosity perhaps one second after
core collapse is in the range of $5\times10^{52}$ ergs s$^{-1}$, 
divided roughly equally between neutrino species. The temperature at
the radius of neutrino decoupling, the neutrinosphere, is $\sim$5 MeV,
so that the pressure scale height is 
$h\approx R_{\rm NS}(R_{\rm NS}/GM)(k_B T/m_p)\simeq0.25$ km
for a PNS radius and mass of $R_{\rm NS}\simeq10$ km and 
$M\simeq1.4$ M$_\odot$, respectively.  
The radius where neutrino heating becomes appreciable, the last point in the
flow where kinetic equilibrium can be maintained, and where the wind-atmosphere 
becomes radiation dominated, is characterized by a specific entropy of $s\approx4$
(see the detailed discussion of Qian \& Woosley 1996).  
Assuming that neutrino absorption on free nucleons dominates heating, this condition determines
a characteristic thermal pressure in the proto-neutron star atmosphere of 
\beq
P_{s=4}\approx3.3\times10^{28}L_{\bar{\nu}_e,52}^{2/3}\,\varepsilon_{\bar{\nu}_e,15}^{4/3}
\,R_{10}^{-4/3} \,\,{\rm ergs \,\,cm^{-3}}, 
\label{p4}
\eeq
where we index the total luminosity of the 
proto-neutron star by the luminosity in $\bar{\nu}_e$ neutrinos and 
where $L_{\bar{\nu}_e,52}=L_{\bar{\nu}_e}/10^{52}$ ergs s$^{-1}$,
$\varepsilon_{\bar{\nu}_e,15}=\varepsilon_{\bar{\nu}_e}/15$ MeV is the average $\bar{\nu}_e$ energy, 
and $R_{10}=R/10$ km.

Although the detailed time evolution of $L_\nu$ and $\varepsilon_\nu$ 
are uncertain, equation (\ref{p4}) shows that as the neutron
star cools and  $L_\nu$ and $\varepsilon_\nu$ decrease, so too does $P_{s=4}$.\footnote{
See, e.g., Pons et al.~(1999) for detailed neutrino luminosity and average energy time profiles.}
Setting $B^2/8\pi$ equal to $P_{s=4}$ yields a criterion for 
the magnetic field strength: 
\beq
B_{s=4}\approx9\times10^{14}
L_{\bar{\nu}_e,52}^{1/3}\varepsilon_{\bar{\nu}_e,15}^{2/3}R_{10}^{-4/3}\,\,\,{\rm G}.
\label{b4}
\eeq
If $B\gtrsim B_{s=4}$, then the magnetic energy density is larger than the 
thermal pressure and it will significantly contribute to or dominate the dynamics of 
the outflow.  This estimate shows that large-scale magnetar-strength magnetic fields 
of $\sim$$10^{15}$ G are required in order to dominate the pressure 
at very early times after explosion.
Essential to the following discussion, for fixed surface magnetic field strength, 
the magnetic field becomes increasingly dominant as the proto-neutron star cools. 
For example, taking $L_\nu\propto t^{-1}$ and $\varepsilon_\nu\propto L_\nu^{1/4}$,
we find that $B_{s=4}^2\propto t^{-1}$; over the few decades of $\tau_{\rm KH}$,
$B_{s=4}^2$ changes appreciably.
Thus, by ``strong magnetic field'' we mean in this paper 
that $B_{\rm NS}\gtrsim B_{s=4}(t)$.\footnote{The $s\approx4$ point is 
generally close enough to the proto-neutron star neutrinosphere that it
is unnecessary at the level of the present discussion to distinguish 
the surface magnetic field strength from $B_{s=4}$.}

The words ``rapid rotation'' can be quantified in several ways.  First,
a neutron star with a millisecond spin period has a reservoir of 
rotational energy of
\beq
E_{\rm rot}\approx2\times10^{52}\,
\mns\,R_{10}^2\,P_1^{-2}\,\,{\rm ergs},
\label{erot}
\eeq
where $P_1$ is the spin period in units of 1 ms.   
Therefore, from the standpoint
of the evolution of the supernova remnant, an initial spin period of 
$P\lesssim5$ ms implies that the rotational energy of the neutron star is 
$E_{\rm rot}\gtrsim E_{SN}\approx10^{51}$ ergs.  That is, if there exists 
a mechanism for spindown of the neutron star on any timescale shorter than 
the remnant age --- some way to efficiently communicate the energy
of rotation to the remnant material --- a spin period of $\lesssim5$ ms
implies that this rotational energy will modify the remnant dynamics at order
unity or larger.

A second estimate of what constitutes ``rapid rotation'', comes from comparing
the spin period with the characteristic convective eddy turn-over timescale 
($\tau_{\rm con}$) within the proto-neutron star during the cooling epoch.  
As argued by Duncan \& Thompson (1992),
the Rossby number ${\cal R}=P/\tau_{\rm con}$ determines whether or not an efficient 
dynamo operates in the proto-neutron star; for $P\sim1$ ms, the Rossby number
is less than unity and conditions are ripe for dynamo action.  Duncan \& Thompson (1992)
argue that it is by virtue of their millisecond rotation at birth that magnetars
develop $10^{14}-10^{15}$ G magnetic fields.  

A last and physically distinct estimate of ``rapid rotation'' comes from 
noting that the breakup spin period of a neutron star is of order 0.5 ms.

In summary, ``strong magnetic field'' means $B\gtrsim10^{14}$ G at early
times after core collapse and the criterion for magnetic field domination
in the proto-neutron star atmosphere {\it decreases} as a function of time.
Additionally, to order of magnitude, ``rapid rotation'' means millisecond
spin periods.\footnote{However, it should be born in mind that both the total 
energy stored in rotation and the properties of the proto-neutron star
atmosphere as breakup is approached are very strong functions of the 
spin period ($P^{-2}$ and exponential in $P^{-2}$, respectively).}
An essential point is that both conditions may be met by {\it all} magnetars at birth.
Indeed, Duncan \& Thompson (1992) argue that $P$ and $B$ are inseparably linked.

\subsection{Millisecond Proto-Magnetars \& \\ Gamma Ray Bursts}

The combination of rapid rotation and high magnetic fields has dramatic
consequences for the dynamics of any outflow that accompanies proto-magnetar
cooling.  As in studies of magnetic winds from rotating 
stars, the winds from proto-magnetars are dominated by magneto-centrifugal 
forces.  Like beads on a wire, the magnetic field lines force the wind material
into corotation with the stellar surface out to the Alfv\'{e}n point, where the 
magnetic energy density equals the kinetic energy density of the outflow. 
This provides an efficient mechanism for spindown
(Schatzman 1962; Weber \& Davis 1967; Mestel 1968ab; Pneuman \& Kopp 1971;
Belcher \& Macgregor 1976; Mestel \& Spruit 1987).  
Thus, a proto-magnetar's rotational energy can be tapped and
communicated, in the form of an energetic outflow, to the surrounding 
medium.  

Although the problem of magnetar birth, and how it differs from the birth
of more typical neutron stars, is interesting in its own right,
there are several reasons for considering millisecond proto-magnetars as the 
central engine of long-duration GRBs: (1) the reservoir of rotation energy is 
in the range required to powers GRBs, (2) proto-magnetar winds,
as for probably all neutron stars, should become relativistic
on the $10-100$ second Kelvin-Helmholtz cooling timescale, and (3) 
the strong observational connection between core-collapse supernovae 
and long-duration GRBs is easy to understand in such a model.  
Of course, because magnetars
are thought to constitute a relatively large fraction of all supernovae, 
not every magnetar can produce a canonical long-duration GRB.
As in the collapsar model for GRBs (Woosley 1993; Macfadyen \& Woosley 1999),
it may be that only millisecond magnetars born within compact Type-Ic
progenitors produce energetic  GRBs with $\sim$$10^{51}$ ergs.
The relativistic outflow from millisecond magnetars probably cannot
sustain high kinetic luminosity for the $\sim$ hours required in 
an extended Type-II progenitor.  Alternatively, it may be that
only those magnetars born with the highest fields and most rapid 
rotation yield conditions favorable for producing a GRB.

Assessing millisecond proto-magnetars  as GRB central engines 
requires answering a number of questions:
(1) What is the spindown timescale, and how much energy is extracted
as a function of time as the proto-magnetar cools? (2) How relativistic is the
outflow and how is the asymptotic Lorentz factor
connected with the energy loss rate? (3) What is the wind geometry,
and specifically, how might that geometry change as a function of
time as the wind transitions from non-relativistic to highly
Poynting-flux dominated?  Finally, (4) how are the relativistic outflow
and the supernova coupled? Are there potential nucleosynthetic
signatures of magnetar spindown? And what might we expect generically from 
the remnants of supernovae that birthed millisecond proto-magnetars?
In this paper we address (1) in detail (\S\ref{section:spindown}),
drawing on analytic estimates and recent exploratory numerical 
models.  We emphasize connections with magnetic stellar winds 
on one hand, and force-free models of pulsar spindown on the other ---
each representing the separate limits in the life of a
millisecond magnetar. We discuss (2), (3), and (4) as well
as the phases of PNS spindown in \S\ref{section:discussion}. 

\section{Millisecond Proto-Magnetar Spindown}
\label{section:spindown}

If the surface magnetic field becomes larger than $B_{s=4}$, either
because $L_\nu(t)$ decreases sufficiently, or $B_{NS}$ increases sufficiently
as a result of dynamo action, the previously solely neutrino-driven outflow 
begins to be accelerated by the action of magneto-centrifugal
forces.  The wind begins to efficiently extract 
rotational energy from the newborn neutron star.
At this stage the outflow is non-relativistic and
the timescale for a millisecond magnetar's spin period to $e$-fold ---
the spindown timescale --- is 
\beq
\tau_{\rm J}=\frac{\Omega}{\dot{\Omega}}\approx
\frac{2}{5}\left(\frac{M}{\dot{M}}\right)\left(\frac{R}{R_{\rm A}}\right)^2,
\label{tauj}
\eeq
where $M$ is the stellar mass, $R$ is the stellar radius, $\dot{M}$ is the mass 
loss rate, and $R_A$ is the radial position of the Alfv\'en point.

Under the assumption that $P\approx1$ ms, 
because $E_{\rm Rot}\gg E_{\rm SN}$ (eq.~\ref{erot}), 
just one $e$-folding of $\Omega$ is sufficient
to modify the dynamics of the supernova remnant significantly
and potentially power a GRB.  
If $\tau_{\rm J}$ is small with respect to the
time for the supernova shock to traverse the progenitor ($\sim$ tens of seconds for type-Ib, -Ic
progenitors) we also expect this extra energy source to modify the supernova nucleosynthesis
(Thompson et al.~2004).

\subsection{The Dominance of Magneto-Centrifugal Forces}

The asymptotic velocity of a wind accelerated predominantly
by magneto-centrifugal forces is $V_\infty\sim R_A\Omega$.  Therefore, a rough
criterion for the dominance of magneto-centrifugal forces  in proto-magnetar winds is
that $R_A\Omega\gtrsim V_\nu$, where $V_\nu$ is the asymptotic velocity
of an outflow accelerated solely by neutrino heating ($\lesssim0.1c$).
For $P\sim1$ ms and $R_{\rm A}\gtrsim15$ km this criterion is 
satisfied, the wind is driven primarily by magneto-centrifugal slinging, neutrino heating 
is relatively unimportant in determining the asymptotic wind velocity,
and rotational energy is transferred 
efficiently from the proto-magnetar to the outflow. 

In order to estimate the spindown timescale and the asymptotic 
velocity of the wind, we must first estimate the Alfv\'en radius.
In what follows, we discuss the non-relativistic limit and then
the transition to the relativistic regime.  We then discuss these
estimates in light of recent numerical models, which more fully
capture the essential physics.

\subsection{The Non-Relativistic Regime:\\ Magnetic Stellar Winds}
\label{section:nonrel}

Angular momentum conservation implies that 
$\dot{J}=d/dt(I\Omega)\approx-\dot{M}R_{\rm A}^2\Omega$.
The location of the Alfv\'en point depends on the radial dependence of the poloidal 
magnetic field. For the purposes of making a simple estimate, we assume
that the field is strictly monopolar so that $B(r)=B_0(R_{NS}/r)^2$.
Using $\rho=\dot{M}/4\pi r^2 v_r$ and the fact that the Alfv\'en speed is 
$v_{\rm A}=v_r(R_{\rm A})\sim v_\phi(R_{\rm A})\sim R_{\rm A}\Omega$, 
the location of $R_A$ is simply
\beqa
R_A&=&B^{2/3}\,R^{4/3}\,(\dot{M}\Omega)^{-1/3}, \nonumber \\
&\sim&40\,\,B_{15}^{2/3}R_{10}^{4/3}\dot{M}_{-3}^{-1/3}P_1^{1/3}\,\,\,{\rm km},
\label{ra}
\eeqa
where $v_r$ is the radial velocity, $v_\phi$
is the azimuthal velocity, $\rho$ is the mass density, 
and $B$ is the radial magnetic field.
In equation (\ref{ra}) $B_{15}=B/10^{15}$ G and $\dot{M}$ is scaled in units of 
$10^{-3}$ M$_\odot$ s$^{-1}$.  

In NRNM protoneutron star winds
the mass loss rate is  (Qian \& Woosley 1996)
\beq
\dot{M}_{\rm NRNM}\approx4\times10^{-5}
L_{\bar{\nu}_e,52}^{5/3}\,\varepsilon_{\bar{\nu}_e,15}^{10/3}
\,R_{10}^{2/3}\,\,{\rm M_\odot\,\,s^{-1}}
\label{mdotnrnm}
\eeq
for a 1.4 M$_\odot$ neutron star,
considerably less than that implied by the scaling of 
equation (\ref{ra}).  When the magnetic field is
strong, centrifugal forces extend the scale height of the
proto-magnetar atmosphere.  Because the mass outflow rate
is essentially determined by the density profile at the sonic
point ($R_s$), this effect can increase $\dot{M}$ significantly 
if $R_A$ is larger than the nominal $R_s$.
For a wind dominated by magneto-centrifugal acceleration, one
finds that 
\beq
R_s\approx(GM/\Omega^2)^{1/3}\approx17\,P_1^{2/3} \,\,{\rm km}.
\label{rs}
\eeq
Because equation (\ref{ra}) implies that $R_A > R_s$ for fiducial 
values of $B$ and $\dot{M}$, the larger scaling is in part justified.
Indeed, more detailed calculations show that for $P=1$ ms and 
$L_{\bar{\nu}_e,52}=1$, $\dot{M}\approx10^{-3}$ M$_\odot$ s$^{-1}$
(Thompson et al.~2004; Metzger et al.~2006).

Using equation (\ref{ra}) we estimate that 
the absolute value of the rotational energy loss rate is 
\beqa
\dot{E}_{\rm NR}&\sim& 
B^{4/3}R^{8/3}\dot{M}^{1/3}\Omega^{4/3} \nonumber \\
&\sim&10^{51}B_{15}^{4/3}R_{10}^{8/3}\mdotth^{1/3} P_1^{-4/3}
\,\,\,{\rm ergs\,\,s^{-1}},
\label{nonreletot1} 
\eeqa
The subscript `NR' is added to emphasize that when the flow is non-relativistic,
$\dot{E}$ depends explicitly on $\dot{M}$.  
The spindown timescale $\Omega/\dot{\Omega}$ in the non-relativistic limit is 
\beq
\tau_{\rm J_{NR}} \simeq 30\,\,{\rm s}\,\,\,
\mns\,\mdotth^{-1/3}\,R_{10}^{-2/3}\, B_{15}^{-4/3}\,P_{1}^{-2/3}.
\label{slowspinscale2} 
\eeq
Note that for slower rotation, larger spin periods, the spindown timescale
{\it decreases} at fixed $\dot{M}$.

There are several uncertainties in these simple estimates.  First, the wind is not
isotropic, so that the factor of $4\pi$ that appears in the relation between
$\dot{M}$ and $\rho$ is incorrect.  Second, the radial velocity, even in the
idealized problem presented here is not $R_A\Omega$, but can differ at the 
factor of two level. Most importantly, the field 
is assumed monopolar, when in reality the surface field must be dipolar
at lowest order.  However, one may argue that even if the star has a dipole
field it is the open field lines that carry the energy and angular momentum,
and perhaps these may be --- to first approximation --- considered monopolar.
For this reason, it turns out that the assumption of 
a monopole field is remarkably good, as long as the surface field 
strength is suitably renormalized to reflect only the field lines
that are opened to inifinity by the wind.  We return to this 
issue below in discussing numerical models.

\subsection{The Non-Relativistic to Relativistic Transition}

As the neutrino luminosity decreases, the characteristic thermal pressure 
in the atmosphere of a proto-magnetar decreases (eq.~\ref{p4}).  For fixed 
surface magnetic field strength, we thus
expect millisecond magnetar winds to become increasingly magnetically-dominated.  
In particular, $\dot{M}$ in equation (\ref{ra}) should decrease as $L_\nu$ decreases
(see eq.~\ref{mdotnrnm}) 
and therefore $R_A$ should increase.   Although $L_\nu$ can decrease arbitrarily 
as $t\rightarrow\infty$,
$R_A$ cannot increase indefinitely; instead, it approaches the radius of the light cylinder
$R_{\rm L}=c/\Omega\simeq48 P_1$ km asymptotically.
As it does so, the flow becomes increasingly relativistic.
This is the transition between non-relativistic magnetically-dominated
mass-loaded outflow and relativistic Poynting-flux dominated wind.  
All neutron stars likely pass through such a transition, 
regardless of their initial spin period and 
magnetic field strength.   Millisecond proto-magnetars are particularly 
interesting as a candidate central engine for GRBs 
because this transition to relativistic flow 
occurs on the Kelvin-Helmholtz timescale ($\sim10-100$\,s) and
at high wind kinetic luminosity, and because magnetars are born in supernovae.

Setting $R_A=R_L$ in equation (\ref{ra}) we can estimate the critical mass loss rate
below which the wind becomes relativistic:
\beqa
\dot{M}_{\rm crit}&=&B^2R^4\Omega^2c^{-3} \nonumber \\
&\sim&7\times10^{-4}B_{15}^2R_{10}^4P^{-2}_{1}\,\,\,{\rm M_\odot\,\,s^{-1}},
\label{mdotcrit}
\eeqa
where $B$ refers to the equivalent surface monopole
field strength.  For lower effective field strengths, $\dot{M}_{\rm crit}$
decreases ---  that is, the mass flux must decrease further in order
to enter the relativistic regime.  For a neutron star born with a 
10 ms spin period and a $10^{12}$ G surface field, 
$\dot{M}_{\rm crit}\sim7\times10^{-12}$ M$_\odot$ s$^{-1}$.
Thus, for weaker fields, the transition occurs at lower $\dot{M}$,
lower $L_\nu$, and at a time longer after collapse and explosion,
a time later in the Kelvin-Helmholtz cooling epoch.

\subsection{Relativistic Winds \& The Force-Free Limit}
\label{section:forcefree}

Once $R_A$ becomes close to $R_L$, 
the degree of Poynting-flux domination is quantified by the parameter (Michel 1969; Goldreich \& Julian 1970)
\beq
\sigma(R_L)=\left. \frac{B^2}{4\pi\gamma\rho c^2}\right|_{R_{\rm L}},
\label{sigma}
\eeq
where $\gamma$ is the Lorentz factor and $B=B(R_L)$. If the outflow is 
driven primarily by magneto-centrifugal forces, $\gamma(R_L)\sim1$.
Roughly speaking, if energy transfer from the electromagnetic field to the 
matter is efficient, $\sigma(R_L)$ measures the maximum acheivable
asymptotic Lorentz factor of the outflow as $r\rightarrow\infty$.  
Assuming that $\gamma(R_L)\sim1$,
$\sigma(R_L)$ can be written simply in terms of the mass loss rate:
\beqa
\sigma(R_L)&\approx& B^2 R_{NS}^{4}\Omega^{2}c^{-3}\dot{M}^{-1} \nonumber \\
&\sim&70\,B_{15}^2R_{10}^4P^{-2}_{1}\dot{M}_{-5}^{-1}
\label{sigmalim}
\eeqa
where $\dot{M}_{-5}=\dot{M}/10^{-5}$ M$_\odot$ s$^{-1}$ and
we have again assumed a monopole field geometry. 
Compare equation (\ref{sigmalim}) with equation (\ref{mdotcrit});
setting $\dot{M}=\dot{M}_{\rm crit}$ in the former, yields $\sigma(R_L)=1$.
Thus, as $L_\nu$ continues to decrease as the cooling epoch progresses,
we expect the flow to become increasingly relativistic and Poynting-flux
dominated ($\sigma$ increases).

In the limiting case of force-free electrodynamics one neglects the
inertia of the matter completely.  This limit corresponds formally
to $\sigma\rightarrow\infty$.  Recent numerical models have calculated 
the spindown luminosity for a neutron star with aligned magnetic
and rotational axes in the force-free limit 
(Contopolous et al.~1999; Gruzinov 2005; Komissarov 2006; Spitkovsky 2006; McKinney 2006); they
find that 
\beqa
\dot{E}_{\rm FF}&=&B^2 R_{NS}^{6}\Omega^{4}c^{-3}\nonumber \\
&\sim& 5.8\times10^{49}B_{15}^{2}R_{10}^{6}P_1^{-4}\,\,{\rm ergs\,\,s^{-1}},
\label{reletot}
\eeqa
where $B$ here corresponds to the surface dipole field strength.  In the 
case of an orthogonal rotator --- with the magnetic axis perpendicular to the
rotation axis --- $\dot{E}_{\rm FF}$ is larger than that in equation (\ref{reletot})
by precisely a factor of two (Spitkovsky 2006).

In the force-free limit, the spindown timescale for the aligned rotator is 
\beq
\tau_{\rm J_{FF}} \simeq 760\,\,{\rm s}\,\,\,
\mns\,R_{10}^{-4}\,B_{15}^{-2}\,P_{1}^{2}.
\label{relspinscale3}
\eeq
It is tempting to compare $\dot{E}_{\rm FF}$ and $\tau_{\rm J_{FF}}$
with equations (\ref{nonreletot1}) and (\ref{slowspinscale2}) directly.
Unfortunately, such a comparison is complicated by the fact that the
former are written assuming a dipole surface field strength, while
the latter assume a monopole field strength.  One requires knowledge
of the amount of open magnetic flux at a given $\dot{M}$ to make
a fair comparison.  As we discuss in more detail below, the results of 
Bucciantini et al.~(2006) show that for typical millisecond proto-magnetar
parameters the equivalent monopole field strength is a factor of $\sim3$
less than the dipole field strength.  Therefore, to compare $\dot{E}_{\rm NR}$
and $\dot{E}_{\rm FF}$, the former should be decreased by a factor of 
$\sim4$, implying that $\dot{E}_{\rm NR}/\dot{E}_{\rm FF}\simeq4
R_{10}^{-10/3}\mdotth^{1/3}P_1^{8/3}$.
This expression has a strong dependence on $P$ and a rather weak dependence
on $\dot{M}$. For example, taking $P=3$ ms one finds that 
the ratio $\dot{E}_{\rm NR}/\dot{E}_{\rm R}$
is nearly 20 times larger.   Thus, the monopole scalings derived in 
\S\ref{section:nonrel} suggest that the early spindown of proto-magnetars is
rapid because the flow is non-relativistic, but magneto-centrifugally driven.  
The spindown luminosity is significantly larger in this phase than an application of 
the force-free spindown law would suggest.  

The original suggestion that
rapidly rotating, highly magnetic neutron stars might power GRBs in 
Usov (1992) essentially employed the force-free approximation.
Similar, but more fully conceived, models were developed by 
Thompson (1994) and Wheeler et al.~(2000).  The latter 
addressed the non-relativistic wind phase, but with a dipole-like
scaling for the magnetic field strength.  Thompson et al.~(2004),
whose analytic estimates we have so far in essence summarized also constructed 
approximate numerical solutions of magneto-centrifugal winds 
from proto-magnetars in order to assess the importance of these 
forces in setting the mass loss rate and the spindown luminosity.

\subsection{Numerical Results}

Two sets of results for the spindown luminosity have been presented in the preceding 
discussion: (1) the non-relativistic, but magneto-centrifugally dominated monopole
(eq.~\ref{nonreletot1}) and (2) the aligned force-free rotator (eq.~\ref{reletot}).  
At face value, the latter is more
secure, coming directly from recent numerical calculations.  However, the applicability
of the force-free result is suspect because it is only formally valid in the limit 
$\sigma\rightarrow\infty$.  We expect quantitative and potentially 
qualitative differences between this limit
and the case where $\sigma$ is merely, say, 2, 10,  or 100. 
We wish to understand the applicability  and limitations of the 
two limits in the millisecond proto-magnetar context.  A set of
recent numerical calculations address these issues directly.

Metzger et al.~(2006) solve the time-dependent one-dimensional 
Weber \& Davis (1967) problem of a non-relativistic magnetically-dominated flow in the 
equatorial plane, including neutrino heating and cooling 
and an appropriate equation of state.  This approach necessarily 
employs a monopole magnetic field configuration.  They
provide a detailed account of the physics of millisecond
proto-magnetar spindown from thermal-magneto-centrifugal winds.  
We emphasize just two results from the study of Metzger et al.~(2006):
(1) because $R_A\Omega$ overestimates the radial velocity at the 
Alfv\'en point,
the spindown timescale in the monopole limit is generally a factor of 
$1.5-2$ {\it lower} than that given in equation (\ref{slowspinscale2})
and (2) the mass loss rate increases exponentially when $R_A$
is larger than the sonic point (eq.~\ref{rs}) so that we do expect an epoch of enhanced
mass loss as the proto-magnetar wind becomes increasingly 
magnetically-dominated, even as $L_\nu$ decreases.

Bucciantini et al.~(2006) (B06) solve the two-dimensional axisymmetric 
time-dependent wind problem with both monopole and dipole surface magnetic fields,
in general relativity.  The mass outflow rate is determined self-consistently
by imposing a finite thermal pressure at the neutron star surface, as 
appropriate in the proto-magnetar context.  They incude the effects of 
neutrino heating and cooling in a parameterized way by employing a 
$\Gamma$-law equation of state. B06 explore
both low magnetization non-relativistic winds and relativistic
Poynting-flux dominated outflows with $\sigma>1$.  For the non-relativistic,
but magnetically-dominated monopole, they find good agreement with 
equation (\ref{nonreletot1}).  In the high-$\sigma$ regime, they 
confirm the analytic force-free monopole limit (Michel 1991; Beskin et al.~1998). 

B06 find very interesting results for the
aligned dipole.  For a strong surface dipole field a region of 
closed magnetic field lines (the ``closed zone'') develops at 
mid-latitudes around the equator in a helmet streamer-like
configuration.  At higher latitudes, the field lines are opened
to infinity by the outflow. The last closed field line meets the
equatorial plane at a $Y$-type point and its radial position
in that plane is denoted $R_Y$.  In the force-free limit one 
expects $R_Y=R_L$ at the equator (e.g., Contopolous et al.~1999).
In the dipole case, there is not a one-to-one correspondence
between the magnitude of the surface magnetic field strength
and the open magnetic flux because as $B$ is increased, the 
closed zone becomes larger.  For this reason, B06
find it convenient to write an approximate relation, derived from 
the simulations, that relates the dipole surface field strength at
the pole to an equivalent monopole surface field strength:
\beq
B_r(R_{NS},\theta=0)\approx B_{r,{\rm eq-m}}(R_{NS})\times\left(\frac{1.6 R_Y}{R_{NS}}\right).
\label{bequiv}
\eeq
B06 find that the spindown luminosity of the aligned dipole with 
surface field $B_r(R_{NS},\theta=0)$ is  the same as that for the
2D monopole if they normalize in terms of the open magnetic flux.
Thus, the aligned dipole with $B_r(R_{NS},\theta=0)$ spins down like 
a monopole with $B_{r,{\rm eq-m}}(R_{NS})$ given by equation (\ref{bequiv}).
Importantly, for $\sigma$ as large as $\sim20$, B06
find that $R_Y$ is considerably less than $R_L$ and that 
the ratio $R_Y/R_L$ is a weak function of the magnetic field
strength.  For parameters typical of millisecond proto-magnetars,
they find that $R_Y/R_L\sim1/4-1/2$.
The direct implication of this result is that the monopole
scalings for spindown derived in \S\ref{section:nonrel}
are applicable to spindown with dipole fields, but that 
the magnetic field strength should be decreased in, for example,
equation (\ref{nonreletot1}) by a factor given by 
equation (\ref{bequiv}).  Thus, if one assumes a dipole field
strength of $10^{15}$ G and that $R_Y/R_{NS}=2$ (compare with B06),
then a field strength of $10^{15}/3.2$ should appear in equation (\ref{nonreletot1}).
Indeed, B06 find that even as $\sigma$ increases to 
values as large as $\sim20$, spindown is {\it more} efficient than 
an application of the pure dipole force-free limit would predict (eq.~\ref{reletot})
because $R_Y$ is significantly less than $R_L$.  These results suggest
that only at significantly higher $\sigma$ does $R_Y$ approach $R_L$
and the force-free limit obtain.  In the context of proto-magnetar spindown,
this means that the transition to the force-free limit occurs at 
smaller $L_\nu(t)$, later in the Kelvin-Helmholtz cooling epoch.

\section{Discussion \& Conclusions: \\ Are Millisecond Proto-Magnetars GRB Central Engines?}
\label{section:discussion}

\subsection{Phases of Spindown Evolution}
\label{section:phase}

There are five phases of spindown in any very young 
rotating neutron star's life (see also B06): (1) a pressure-dominated phase in
which the wind is driven by neutrino-heating ($B<B_{s=4}$; eq.~\ref{b4}), 
(2) a phase in which magnetic field effects are present,
but not dominant so that $R_A\Omega<0.1c\approx c_T$,
where $c_T$ is the isothermal sound speed at the proto-neutron star surface,
(3) a non-relativistic magnetically-dominated phase when 
$R_A$ is greater than $R$ and $R_s$, but less than $R_L$, (4)
a relativistic phase in which $R_A\sim R_L$, but $R_Y<R_L$ (as in B06),
and lastly (5) an epoch when the force-free limit is 
(presumably) applicable and $R_Y\simeq R_A\simeq R_L$.
Roughly speaking, for monotonically decreasing $L_\nu$, 
phases (1)$-$(5) represent a time evolution starting immediately
after the supernova explosion commences.

The transition from phase (1) to phase (2) is dictated by equation
(\ref{p4}).  When $B^2/8\pi$ exceeds $P_{s=4}$ near the proto-magnetar
surface, phase (2) begins.  As the magnetic field becomes increasingly
important, the sonic point will move to smaller radii and the radial scale
over which the magnetic field dominates will move to larger radii.  
When $R_A$ is of order $R_s$, phase (3) begins.  This transition
will be complicated by the fact that $\dot{M}$ may increase
as a result of increased magneto-centrifugal support in the 
quasi-hydrostatic atmosphere (see Metzger et al.~2006 for details).
Throughout phase (3) $R_A\Omega>c_T$,
$\sigma<1$, and $R_A$ increases
from of order the slow magnetosonic radius to $R_L$.  
The characteristic spindown luminosity is 
\begin{equation}
\dot{E}\approx 4\times10^{50}\, 
B_{14.5}^2 R_{10}^4 P_{1}^{-5/3} M_{1.4}^{-1/3}\,\,\,{\rm ergs\,\,s^{-1}},
\label{energyeq}
\end{equation}
where $B$ is the equivalent monopole field (see eq.~\ref{bequiv}).
Because we expect this phase to last of order the Kelvin-Helmholtz
timescale, the total amount of energy extracted is comparable to the
asymptotic supernova energy, $\sim10^{51}$ ergs.

The transition from phase (3)
to phase (4) occurs at $\sigma\approx1$ ($R_A\approx R_L$) and 
for a critical $\dot{M}$ given by equation (\ref{mdotcrit}).
Throughout phase (4) the flow is relativistic and Poynting-flux dominated ($\sigma>1$).
The total energy and angular momentum loss rates are smaller than in phase (3).  
However, because $R_Y<R_L$, the spindown rate is larger than what would
be inferred from an application of the force-free limit.  
The results of B06 indicate that this is true despite
the fact that $\sigma$ is larger than and, increasingly with time as the 
neutrino luminosity decreases, much larger than unity.
Eventually, $L_\nu$ and $\dot{M}$ decrease sufficiently that 
$R_Y\approx R_L$, and phase (5) begins. A simple and very uncertain
extrapolation of the results of B06 suggest that this transition occurs at 
a very high $\sigma$ (perhaps $\sim10^6$).  In this epoch, the spindown luminosity
is given by the recent force-free calculations described in \S\ref{section:forcefree}.

There is a last phase (or set of phases) beyond the scope of the present work,
which follow after the Kelvin-Helmholtz cooling epoch, as the MHD approximation
in the magnetosphere of any young neutron star or magnetar begins to break down,
the flow becomes charge-separated, and particles are accelerated electromagnetically
and to high Lorentz factors directly off of the neutron star surface.

\subsection{Energetics, Relativity, \& Variability}

As first emphasized by Usov (1992), the energy budget and luminosity
of millisecond magnetar spindown 
is in the range needed to explain cosmological long-duration
GRBs (eqs.~\ref{erot} \& \ref{reletot}; see also Thompson 1994; Wheeler et al.~2000;
similar models also by  Katz 1997; 
Kluzniak \& Ruderman 1998).  
The recent work of Thompson et al.~(2004), Metzger et al.~(2006), and B06,
which we have reviewed here, have explored a more complete picture of 
proto-magnetar spindown, from the initial non-relativistic wind stage through the 
Kelvin-Helmholtz cooling epoch.  

In order to power a GRB, the outflow
must simultaneously achieve high spindown luminosity and high Lorentz factor
($\gamma_\infty\gtrsim100$; e.g., Lithwick \& Sari 2001).  The asymptotic
Lorentz factor depends on the magnetization of the flow,
measured by $\sigma$ (eq.~\ref{sigma}), which is, in turn,
set by the time dependence of $\dot{M}$ (eq.~\ref{sigmalim}), itself determined by 
$L_\nu(t)$.  Thus, the transition to a relativistic flow ($\sigma>1$)
is governed by the Kelvin-Helmholtz cooling timescale $\tau_{\rm KH}\sim10-100$s.  
If we take the average duration of a long GRB to be $\tau_{\rm GRB}\sim30$s, the constraints
that (1) the total energy ejected must be of order $E_{\rm GRB}\sim10^{51.5}$ ergs and (2)
that $\gamma_\infty\sim100$ imply that there must be efficient conversion of 
electromagnetic energy to kinetic energy in the outflow beyond $R_L$.
The reason follows from noting that $\sigma\propto\dot{M}^{-1}$ and 
that $\dot{E}$ is a decreasing function of $\dot{M}$ before the force-free
limit is reached, after which $\dot{E}$ is independent of $\dot{M}$.  
These facts limit the space of possible $\sigma$ obtainable at a given 
$\dot{E}$ to the range of several hundred, depending on $B$ and $\Omega$
at the proto-magnetar surface.  Therefore, in order to have the 
Lorentz factors required for GRBs, $\gamma_\infty\sim\sigma(R_L)$.
This is in contradiction to expectations from the force-free monopole,
which gives $\gamma_\infty\sim\sigma(R_L)^{1/3}$ and 
is potentially accomplished by efficient
magnetic dissipation in the out-flowing wind at radii larger than or comparable to 
$R_L$ and/or the fast magnetosonic point (Drenkhahn \& Spruit 2002; Lyutikov \& Blandford 2003).

We expect variability in both $\dot{E}$ and $\sigma$ due to rapid
changes in mass loading.  In an average sense, this may cause the wind
to alternate rapidly between $\sigma<1$ and $\sigma>1$.
Proto-magnetars, like all neutron stars,
are expected to be fully convective during $\tau_{\rm KH}$
(Keil et al.~1996; Pons et al.~1999; Thompson \& Murray 2001;
Dessart et al.~2006a).  Large-scale
convection, magnetic field footpoint motion, and instabilities in the
magnetosphere may all cause variability in these important quantities ($\dot{E}$, $\gamma$, $\sigma$)
on millisecond timescales.   Strong variations in the mass loading 
could be caused by shearing of large-scale closed magnetic loops on the surface of the
fully convective millisecond proto-magnetar core (Thompson 1994).
Internal shocks in the flow are a natural consequence of 
these rapid changes in $\dot{M}$ (Rees \& M\'{e}sz\'{a}ros 1994).

\subsection{Emergence \& Geometry}

Evidence for collimation abounds in afterglow observations of GRBs.  
On the other hand, there is much theoretical work 
supporting the conclusion that it is difficult to collimate Poynting-flux 
dominated flows (e.g., Begelman \& Li 1994; Lyubarsky \& Eichler 2001).  
If the relativistic proto-magnetar wind can be collimated, 
we expect its emergence from the progenitor
to resemble models of collapsar jets escaping their Type-Ibc hosts 
(e.g., Aloy et al.~2000; Zhang et al.~2003, 2004).  But, can relativistic
proto-magnetar winds be collimated?  

There are a number of possible answers.  The first potential answer may be 
that collimation simply cannot be accounted for in the millisecond proto-magnetar model for GRBs. 
The second reply is that relativistic Poynting-flux dominated winds actually {\it can}
be efficiently collimated (see Vlahakis \& K\"onigl 2003; Vlahakis 2004).  A third possibility 
is that a small disk forms outside the rapidly rotating magnetar 
and that this aids collimation, or that the proto-magnetar is so distorted
by centrifugal forces that it is disk-like at early times (e.g., see Figs.~7 \& 8 of 
Dessart et al.~2006b).  This option might provide a picture which connects 
logically with the collapsar model of GRBs (Macfadyen \& Woosley 1999).
A fourth option, as suggested by Thompson (2005), is to appeal to a somewhat different
geometry.  Pulsars drive energetically dominant high Lorentz
factor {\it equatorial} outflows (Komissarov \& Lyubarsky 2004; Spitkovsky \& Arons 2004).  
The models of B06 show that when $\sigma>1$,
$\gamma(\theta)$ is peaked in the equatorial plane.  In addition,
the presence of the current sheet in this region may facilitate the magnetic
dissipation required for efficient conversion of magnetic energy to kinetic energy.
In analogy with pulsar winds, perhaps it is possible that the geometry of many GRBs
is ``sheet''- or ``fan''-like rather than jet-like so that the solid angle subtended by the
GRB is $\sim\theta$ instead of $\sim\theta^2$.
Modeling shows that the expected
sheet-break (the analog of the jet-break; e.g., Rhoads 1999; Sari et al.~1999) is not steep enough
to explain all achromatic breaks in GRB afterglow 
lightcurves (T.~Thompson, unpublished; Granot 2005).  
In addition, for the same observed isotropic equivalent energy and break time, 
a sheet-like geometry increases the true energy of the burst with respect to 
a uniform jet by a factor $\propto\theta^{-1}$.  Nevertheless,
this geometry remains an interesting alternative for a subset of bursts with shallower
break profiles (see, e.g., Fig.~2 of Panaitescu 2005).

There is a another interesting possibility that may bear on the question
of collimation in millisecond proto-magnetar winds.
The relativistic Poynting-flux dominated wind of phases (4) and (5) comes {\it after} 
the non-relativistic mass-loaded $\sigma<1$ wind of phase (3) (see \S\ref{section:phase}).  
This means that the envelope structure that the relativistic wind encounters
has been ``pre-processed'' by the preceding non-relativistic flow.
The results of B06 (in analogy with models of non-relativistic stellar winds; e.g., Smith 1998) 
show that the energetic flux is strongly directed along the rotation axis by hoop stress
when $\sigma<1$.  Thus, if the total 
energy extracted from the proto-magnetar in phase (3) is on the order of the supernova 
explosion energy, then there will be a relatively ``hollow,'' asymmetric, and elongated channel
that the subsequent relativistic flow emerges into.  Pressure forces and wind material 
bounding this channel may force the less-energetic and relativistic flow into 
a jet-like structure.

\subsection{Remnants \& Nucleosynthesis}

For fiducial proto-magnetar parameters,
the timescale for extraction of an amount of rotational energy comparable
to $10^{51}$ ergs is small on the timescale for the supernova shockwave to 
traverse a Type-II supernova progenitor and comparable to that for a Type-Ibc
progenitor.  Much of this energy is extracted during phase (3) ($\sigma<1$), when the 
energetic flux is strongly directed along the axis of rotation because of 
hoop stress (as in B06).   The action of this outflow may cause an asymmetry 
in the supernova remnant (see also Wheeler et al.~2000).  B06
show that the zenith angle at
which the energetic flux is maximized is an increasing function of $\sigma$
so that for $\sigma>1$, the energetic loss is primarily equatorial.
Depending on the timing of the start of phase (3) with respect
to the position and energy of the preceding supernova shockwave, 
the action of the energetic wind could modify the nucleosynthesis
in the remnant in an asymmetric way.

As a bit of speculation and for illustrative purposes, we note that the 
Cass A supernova remnant has a strong jet/counter-jet morphology with
a distinctive nucleosynthetic enrichment signature (Hwang et al.~2001, 2004;
Willingale et al.~2002; Fesen et al.~2006) and that it has been at least
discussed in the context of GRB remnants (Laming et al.~2006).  Indeed, many
core-collapse supernova events exhibit asymmetry in their 
spectropolarimetry (e.g., Wang et al.~2001).  However, for 
large-scale asymmetries representing total energy comparable to the asymptotic
supernova remnant energetics (as in Cass A), we require surface
magnetic fields of $\gtrsim10^{14}$ G and rotation rates at birth in the 
millisecond range.  Interestingly, Chakrabarty et al.~(2001) suggest that the X-ray point
source in the Cass A remnant is an Anomalous X-ray Pulsar, a magnetar,
and so perhaps this object satisfied our requirements at birth.
Evidence also exists for infrared light echoes from the X-ray point
source, which are interpreted as burst-like events, potentially 
analogous to the giant flares seen from magnetars (Krause et al.~2005; as in, e.g., SGR 1806-20, Palmer et al.~2005). 

Because $\gtrsim10^{51}$ ergs
can be extracted on a timescale shorter than or comparable to the timescale for 
the supernova shockwave to traverse the progenitor, we expect that the wind may 
significantly affect the nucleosynthetic yield and its angular distribution. 
If significant energy can be extracted and communicated to the surrounding
envelope of expanding supernova shocked gas rapidly, the $^{56}$Ni yield of 
proto-magnetars may be enhanced.
In this way it may be possible to generate hyper-energetic or 1998bw-like supernovae
(Thompson et al.~2004; Woosley \& Heger 2003).  
The inferred energetics and $^{56}$Ni yield of SN2003dh
and SN1998bw put strong constraints on any GRB mechanism.  In the collapsar model
a disk wind is thought to generate the $^{56}$Ni required to power the SN lightcurve
(Macfadyen \& Woosley 1999; Pruet et al.~2004).  In the millisecond proto-magnetar model, the energetic
wind shocks the material already processed by the supernova shock, perhaps generating
the large inferred $^{56}$Ni yields.  Such a mechanism relies on timing.  We are currently
investigating this scenario more fully.

\acknowledgements

I thank Niccolo Bucciantini, 
Brian Metzger, Philip Chang, Eliot Quataert, 
and Jon Arons for stimulating conversations 
and collaboration.


\begin{thebibliography}


\bibitem[Aloy et al.~(2000)]{aloy}
Aloy, M.-A., M\"{u}ller, E., Ib\'{a}\~{n}ez, J., Marti, J., MacFadyen, A.~2000, ApJL, 531, L119

\bibitem[Arons (2003)]{arons}
Arons, J.~2003, ApJ, 589, 871

\bibitem[Begelman \& Li (1994)]{begelman_li}
Begelman, M.~C. \& Li, Z.-Y.~1994, ApJ, 426, 269

\bibitem[Belcher \& Macgregor (1976)]{belcher}
Belcher, J.~W.~\& Macgregor, K.~B.~1976, ApJ, 210, 498

\bibitem[Beskin et al.~(1998)]{beskin}
Beskin, V.~S., Kuznetsova, I.~V., \& Rafikov, R.~R.~1998, MNRAS, 299, 341

\bibitem[Bucciantini et al.~(2006)]{bucciantini}
Bucciantini, N., Thompson, T.~A., Arons, J., Quataert, E., Del Zanna, L.~2006,
MNRAS, in press (B06)

\bibitem[Burrows, Hayes, \& Fryxell~(1995)]{bhf1995}
Burrows, A., Hayes, J., \& Fryxell, B. A. 1995, ApJ, 450, 830 

\bibitem[Burrows \& Lattimer (1986)]{bl1986}
Burrows, A. \& Lattimer, J. M. 1986, ApJ, 307, 178

\bibitem[Cardall \& Fuller~(1997)]{cardall} 
Cardall, C. Y. \& Fuller, G. M. 1997, ApJL, 486, 111 

\bibitem[Chakrabarty et al.(2001)]{2001ApJ...548..800C} Chakrabarty, D., 
Pivovaroff, M.~J., Hernquist, L.~E., Heyl, J.~S., \& Narayan, R.\ 2001, 
\apj, 548, 800 

\bibitem[Contopolous et al.~(1999)]{contopolous}
Contopolous, I., Kazanas, D., Fendt, C.~1999, ApJ, 511, 351

\bibitem[Dessart et al.(2006)]{2006ApJ...645..534D} Dessart, L., Burrows, 
A., Livne, E., \& Ott, C.~D.\ 2006a, \apj, 645, 534 

\bibitem[Dessart et al.(2006)]{2006ApJ...644.1063D} Dessart, L., et al.\ 2006b, \apj, 644, 
1063 

\bibitem[Drenkhahn \& Spruit (2002)]{drenkhahn_spruit}
Drenkhahn, G.~\& Spruit, H.~C.~2002, A\&A, 391, 1141

\bibitem[Duncan, Shapiro, \& Wasserman (1986)]{dsw}
Duncan, R. C., Shapiro, S. L., \& Wasserman, I. 1986, ApJ, 309, 141 

\bibitem[Duncan \& Thompson (1992)]{dunc1992}
Duncan, R.~C.~\& Thompson, C.~1992, ApJL, 392, 9

\bibitem[Fesen et al.(2006)]{2006astro.ph..3371F} Fesen, R.~A., et al.\ 
2006, ApJ, in press, arXiv:astro-ph/0603371 

\bibitem[Galama et al.~(1998)]{galama}
Galama, T.~J., et al.~1998, Nature, 395, 670

\bibitem[Goldreich \& Julian (1970)]{goldreich}
Goldreich, P.~\& Julian, W.~H.~1970, ApJ, 160, 971

\bibitem[Granot(2005)]{2005ApJ...631.1022G} Granot, J.\ 2005, \apj, 631, 1022
 
\bibitem[Gruzinov(2005)]{2005PhRvL..94b1101G} Gruzinov, A.\ 2005, Physical 
Review Letters, 94, 021101 

\bibitem[Hjorth et al.~(2003)]{hjorth}
Hjorth, J., et al.~2003, Nature, 423, 847 

\bibitem[Hwang et al.(2001)]{2001ApJ...560L.175H} Hwang, U., et al.\ 2001, \apjl, 560, L175 

\bibitem[Hwang et al.(2004)]{2004ApJ...615L.117H} Hwang, U., et al.\ 2004, 
\apjl, 615, L117 

\bibitem[Kaspi \& Helfand(2002)]{2002ASPC..271....3K} Kaspi, V.~M., \& 
Helfand, D.~J.\ 2002, ASP Conf.~Ser.~271: Neutron Stars in Supernova 
Remnants, 271, 3 

\bibitem[Katz(1997)]{1997ApJ...490..633K} Katz, J.~I.\ 1997, \apj, 490, 633 

\bibitem[Keil, Janka, \& M\"{u}ller (1996)]{keil_janka}
Keil, W., Janka, H.-Th., \& M\"{u}ller, E.~1996, ApJL, 473, L111

\bibitem[Klu{\'z}niak \& Ruderman(1998)]{1998ApJ...505L.113K} Klu{\'z}niak, 
W., \& Ruderman, M.\ 1998, \apjl, 505, L113 

\bibitem[Komissarov \& Lyubarsky (2004)]{komissarov}
Komissarov, S.~S.~\& Lyubarsky, Y.~2004, MNRAS, 349, 779

\bibitem[Komissarov(2006)]{2006MNRAS.367...19K} Komissarov, S.~S.\ 2006, 
\mnras, 367, 19 

\bibitem[Kouveliotou et al.~(1999)]{kouv} Kouveliotou, C., et al.~1999, ApJL, 510, 115

 \bibitem[Krause et al.(2005)]{2005Sci...308.1604K} Krause, O., et al.\ 
2005, Science, 308, 1604 

\bibitem[Laming et al.(2006)]{2006astro.ph..3434L} Laming, J.~M., et al.\ 2006, accepted to ApJ, arXiv:astro-ph/0603434 

\bibitem[Lithwick \& Sari(2001)]{2001ApJ...555..540L} Lithwick, Y., \& 
Sari, R.\ 2001, \apj, 555, 540 

\bibitem[Lyubarsky \& Eichler (2001)]{lyubarsky}
Lyubarsky, Y.~\& Eichler, D.~2001, ApJ, 562, 494

\bibitem[Lyutikov \& Blandford (2003)]{lyutikov}
Lyutikov, M.~\& Blandford, R.~2003, astro-ph/0312347

\bibitem[Maeda et al.~(2003)]{maeda}
Maeda, K., et al.~2003, ApJ, 593, 931

\bibitem[MacFadyen \& Woosley (1999)]{macfadyen} 
MacFadyen, A.~I. \& Woosley, S.~E. 1999, ApJ, 524, 262

\bibitem[McKinney (2006)]{mckinney}
McKinney, J.~C.~2006, accepted for publication in MNRAS Lett., astro-ph/0601411 

\bibitem[Mestel (1968)]{mestel}
Mestel, L.~1968a, MNRAS, 138, 359

\bibitem[Mestel (1968)]{mestel2}
Mestel, L.~1968b, MNRAS, 140, 177

\bibitem[Mestel \& Spruit (1987)]{mestel_spruit}
Mestel, L.~\& Spruit, H.~C.~1987, MNRAS, 226, 57

\bibitem[Metzger et al.(2006)]{2006astro.ph..8682M} Metzger, B.~D., 
Thompson, T.~A., \& Quataert, E.\ 2006, arXiv:astro-ph/0608682 

\bibitem[Michel (1969)]{michel69}
Michel, F.~C.~1969, ApJ, 158, 727

\bibitem[Michel (1991)]{michel}
Michel, F.~C.~1991, Theory of Neutron Star Magnetospheres, Univ.~Chicago Press, Chicago

\bibitem[Otsuki et al.~(2000)]{otsuki}
Otsuki, K., Tagoshi, H., Kajino, T., \& Wanajo, S.-Y. 2000, ApJ, 533, 424

\bibitem[Palmer et al.(2005)]{2005Natur.434.1107P} Palmer, D.~M., et al.\ 
2005, \nat, 434, 1107 

\bibitem[Panaitescu(2005)]{2005MNRAS.362..921P} Panaitescu, A.\ 2005, 
\mnras, 362, 921 

\bibitem[Pneuman \& Kopp (1971)]{pneuman}
Pneuman, G.~W.~\& Kopp, R.~A.~1971, 18, 258

\bibitem[Pons et al.~(1999)]{pons1999}
Pons, J. A., Reddy, S., Prakash, M., Lattimer, J. M., \& Miralles, J. A. 1999, ApJ, 513, 780

\bibitem[Pruet et al.(2004)]{2004ApJ...606.1006P} Pruet, J., Thompson, 
T.~A., \& Hoffman, R.~D.\ 2004, \apj, 606, 1006 

\bibitem[Qian \& Woosley (1996)]{qw1996}
Qian, Y.-Z. \& Woosley, S. E. 1996, ApJ, 471, 331 

\bibitem[Rees \& M\'{e}sz\'{a}ros (1994)]{rees_meszaros1994}
Rees, M.~\& M\'{e}sz\'{a}ros P.~1994, ApJL, 430, 93

\bibitem[Rhoads(1999)]{1999ApJ...525..737R} Rhoads, J.~E.\ 1999, \apj, 525, 737 

\bibitem[Sari et al.(1999)]{1999ApJ...519L..17S} Sari, R., Piran, T., \& 
Halpern, J.~P.\ 1999, \apjl, 519, L17 

\bibitem[Schatzman (1962)]{schatzman}
Schatzman, E.~1962, Annales d'Astrophysique, 25, 18

\bibitem[Scheck et al.(2006)]{2006A&A...457..963S} Scheck, L., Kifonidis, 
K., Janka, H.-T., M\"uller, E.\ 2006, \aap, 457, 963 

\bibitem[Smith (1998)]{smith}
Smith, M.~D.~1998, Ap\&SS, 261, 169

\bibitem[Spitkovsky \& Arons (2004)]{spitkovsky1}
Spitkovsky, A.~\& Arons, J.~2004, ApJ, 603, 669

\bibitem[Spitkovsky(2006)]{2006ApJ...648L..51S} Spitkovsky, A.\ 2006, 
\apjl, 648, L51 


\bibitem[Spruit, Daigne, \& Drenkhahn (2001)]{spruit_daigne_drenk}
Spruit, H.~C., Daigne, F., \& Drenkhahn, G.~2001, A\&A, 369, 694

\bibitem[Stanek et al.~(2003)]{stanek}
Stanek, K.~Z., et al.~2003, ApJL, 591, L17

\bibitem[Sumiyoshi et al.~(2000)]{sumiyoshi} 
Sumiyoshi, K., Suzuki, H., Otsuki, K., Teresawa, M., \& Yamada, S. 2000, PASJ, 52, 601

\bibitem[Takahashi, Witti, \& Janka (1994)]{twj1994} 
Takahashi, K., Witti, J., \& Janka, H.-T. 1994, A\&A, 286, 857 

\bibitem[Thompson (1994)]{thompson1994}
Thompson, C.~1994, MNRAS, 270, 480

\bibitem[Thompson \& Duncan (1993)]{thomp_dunc1993}
Thompson, C.~\& Duncan, R.~C.~1993, ApJ, 408, 194

\bibitem[Thompson \& Murray (2001)]{thompson_murray}
Thompson, C.~\& Murray, N.~2001, ApJ, 560, 339

\bibitem[Thompson et al.~(2002)]{thompson2002}
Thompson, C., Lyutikov, M., \& Kulkarni, S.~R.~2002, ApJ, 574, 332

\bibitem[Thompson, Burrows, \& Meyer (2001)]{thompson1}
Thompson, T.~A., Burrows, A., \& Meyer, B. S.~2001, ApJ, 562, 887   

\bibitem[Thompson (2003a)]{btb}
Thompson, T.~A.~2003a, {\it Core-Collapse of Massive Stars},
Edited by C.~Fryer, Kluwer Academic Publishers

\bibitem[Thompson (2003b)]{thompson3}
Thompson, T.~A.~2003b, ApJL, 585, L33

\bibitem[Thompson et al.~(2004)]{tcq}
Thompson, T.~A., Chang, P., \& Quataert, E.~2004, ApJ, 611, 380

\bibitem[Usov (1992)]{usov_nature}
Usov, V.~1992, Nature, 357, 472

\bibitem[Vlahakis \& K\"onigl (2003)]{vlahakis1}
Vlahakis, N.~\& K\"onigl, A.~2003, ApJ, 596, 1080

\bibitem[Vlahakis (2004)]{vlahakis2}
Vlahakis, N.~2004, ApJ, 600, 324

\bibitem[Wanajo et al.~(2000)]{wanajo} 
Wanajo, S., Kajino, T., Mathews, G. J., \& Otsuki, K. 2001, ApJ, 554, 578

\bibitem[Wang et al.(2001)]{2001ApJ...550.1030W} Wang, L., Howell, D.~A., 
H{\"o}flich, P., \& Wheeler, J.~C.\ 2001, \apj, 550, 1030 

\bibitem[Weber \& Davis (1967)]{weber}
Weber, E.~J.~\& Davis, L.~1967, ApJ, 148, 217

\bibitem[Wheeler et al.~(2000)]{wheeler}
Wheeler, J.~C., Yi, I., H\"{o}flich, P., \& Wang, L.~2000, ApJ, 537, 810

\bibitem[Willingale et al.(2002)]{2002A&A...381.1039W} Willingale, R., 
Bleeker, J.~A.~M., van der Heyden, K.~J., Kaastra, J.~S., \& Vink, J.\ 
2002, \aap, 381, 1039 

\bibitem[Woods \& Thompson(2004)]{2004astro.ph..6133W} Woods, P.~M., \& 
Thompson, C.\ 2004, arXiv:astro-ph/0406133 

\bibitem[Woosley (1993)]{woosley1993}
Woosley, S. E.~1993, ApJ, 405, 273

\bibitem[Woosley et al.~(1994)]{woosley1994}
Woosley, S. E., Wilson, J. R., Mathews G. J., Hoffman, R. D., \& Meyer, B. S. 1994, ApJ, 433, 209

\bibitem[Woosley \& Heger (2003)]{woosley_heger}
Woosley, S.~E.~\& Heger, A.~2003, submitted to ApJ, astro-ph/0301965

\bibitem[Zhang, Woosley, \& MacFadyen  (2003)]{zhang} 
Zhang, W., Woosley, S.~E., \& MacFadyen, A.~I.~2003, ApJ, 586, 356
\end{thebibliography}
\end{document}